# Intranational Skill-relevance Model of the Immigrant's Self-selection:

## Further Evidence of the Stylized Fact from the E-9 Employment Permit System (EPS)

Kwonhyung Lee[*]

*Department of Public Administration, Yonsei University, Seoul 03722, Korea*

Yejin Lim[†]

*International Studies Major, Underwood Int'l College, Yonsei University, Seoul 03722, Korea*

Sunghyun Cho[‡]

*School of Industrial and Management Engineering, Korea University, Seoul 02841, Korea*

### ABSTRACT

This study expands upon the foundation of 'Skill-Relevance-Self Selection' model on labor immigration, introduced by our previous study (Lee, Lim, & Cho, 2022). In detail, we seek an empirical confirmation of the model by providing evidence of the attained -however, yet to be tested- stylized fact: 'as the discount of intranational skill-relevance by immigration is intensified, the wage differential of international labor immigration is diminished'. Utilizing the hypothesis and data meticulously formulated and selected in consideration of Employment Permit System (EPS) and the typology of host nations, OLS linear regression results reasonably support all hypotheses with statistical significance, thereby inductively substantiating our constructed model. This paper contributes to existing labor immigration economics literature in three following aspects: (1) Acknowledge the previously overlooked factor of 'skill relevance discount' in labor immigration as an independent parameter, separate from the 'Moving cost' of Borjas model (1987; 1991); (2) Demonstrate the down-to-earth economic implications of host nation typology, thereby establishing a taxonomy in existence, rather than an ideal classification; (3) Seek a complementary synthesis of two grand strands of research methodology – that is, deductive mathematical modeling and inductive statistical testing.

## 1 Introduction

Faced with ever-low fertility rate and unprecedently fast population aging, Republic of Korea's acceptance of migrant laborers has become something more of a necessity than a choice. Even so, insufficient -or worse, misplaced- conceptions of immigration dynamics hinder the achievement of the intended policy goal; of which in some cases, reach to the policy effect in the very opposite direction. Combined with the patchwork-like interorganizational structure of Korean immigration policies (Lee, Kim, & Rhee, 2022), such misconceptions can be quite detrimental – which calls for a multidimensional, comprehensive analysis into labor immigration economics.

Labor immigration environment of Korea is distant from that of United States; for that the absolute majority of inbound labor immigrants seeking employment in Korea are short-term visitors, rather than long-term residents (Kim, 2020). For this reason, Employment Permit System (EPS) of E-9 Visa, a short-run employer (i.e. labor demander)-centered system that matches job-seeking migrants with small-to-medium firms experiencing excess labor demand, takes the majority of annual budget allocated to labor immigration policy. The concerned migrant laborers of EPS usually make up for excess labor demand of the 'secondary industry', which includes small-to-medium manufacturing (e.g. labor-intensive light industry, simple machinery), construction, mining, and so on.

---

[*] sangsankwon@yonsei.ac.kr

[†] yejinlim@yonsei.ac.kr

[‡] cshhello99@korea.ac.kr

In labor immigration, countries of origin are mostly consisted of developing nations and newly industrializing economies, whereas host nations are mostly advanced economies, with a majority of its native population working in the 'tertiary industry'. Subsequently, in terms of intranational, or country-specific skill relevance, occupational skills specialized in the secondary industry are relatively more appreciated in sending-nations; on the flip side, depreciated in host countries. As such, labor immigrants must strike an optimal balance between the wage differential of immigration (which amplifies their utility) and the 'discount of intranational skill-relevance' entailed by it (which diminishes their utility). Likewise, the firms hiring them should choose the amount of wage accordingly; that is, just to the point where the potential labor immigrants barely decide to accept the labor-wage contract – any additional wage above the point would be irrational for the profit maximization.[1]

Based on the institutional process of EPS, we (Lee, Lim, & Cho, 2022) formerly modelized an archetypal sequential contract of short-term labor immigration between migrants (employees) and firms (employers) – with a unique key parameter formerly overlooked by other economists, namely, 'discount of intranational skill-relevance ($\delta$)' caused by immigration. Utilizing game theory's 'backward induction', the study provided the optimal solution of the very model, that is, Subgame Perfect Nash Equilibrium (SPNE). Afterwards, empirical tests into the economic implications attained by the SPNE were implemented by us; though we conducted regressions controlling effects of either sending-nations or host-nations, statistical research directly addressing the dynamics around the '$\delta$', however, is yet to be made – in which is the very remnant this study seeks to provide an explanation.

## 2 Theoretical Background

### 2.1 *Labor Immigration Policy*

#### 2.1.1 *Host Nation Typology*

Immigration policies are inseparable from a nation's historical, ethnic and cultural background; such underlying contexts include colonial experience, famines, civil wars and international conflicts – which often becomes an obstacle against the attempts seeking formulation of a general theory. Nonetheless, immigration theorists have formed a host (or accepting) nation typology of immigration over the years; such sorting enables a generalization within its respective type, therefore, provides a degree of predictability. The typology in concordance consists of three kinds: traditional, pioneer, and latecomer.

Traditional host nations are countries established through -and built by- immigration (Okólski, 2012a; 2012b); United States, Canada, and Australia are exemplary. Pioneer host nations are defined as countries that have experienced labor supply shortage from the postwar period of WWII to the 1970s (Massey, 1999); western Europe (e.g. United Kingdom, France, (West) Germany) and northern Europe (e.g. Sweden, Norway, Finland) fall in this category. Lastly, latecomer host nations refer to countries that went through 'immigration change', that is, from the status of a 'sending nation' to 'host nation' - since the 1980s; southern Europe (e.g. Italy, Greece, Spain) and advanced economies in Asia (e.g. Korea, Hong Kong, Singapore) are representative cases (Lee, Kim, & Rhee, 2022; Lee, 2010).

In line with their immigration history and institutional heritage, long-term (or permanent) residence, or naturalization would take place more in traditional type nations, but less in latecomer type countries. On the other hand, short-term visitors would be the prevalent type of labor immigration in latecomer countries – which would not be the case in traditional type nations. Pioneer host nations would be in someplace between the two; in regard to economic significance of short-term (or long-term) labor immigration, the three types could be thought of as forming a continuum.

---

[1] The efficiency wage theory/system by Akerlof & Yellen (1988) are out of the question in this context.

**2.1.2** *Korean Context*

Studies into labor immigration policy of Korea can be divided into two main subjects: EPS and immigrants themselves. In studies addressing EPS, approaches based on public administration (e.g. Han, 2022; Syifa; 2017), policy studies (e.g. Woo, 2022; Dyrberg, 2015) and jurisprudence (e.g. Kim, 2021; Yoo, 2019; Choi, 2018) were dominant; whereas vast majority of immigrant-studies were founded on sociology (e.g. Jang, 2022; Callinan, 2020; Seol, 2009) and cultural-anthropology (e.g. Einhorn, 2022; Kim, 2022; Wigglesworth & Fonseca, 2016). However, in both subjects, there were relatively less studies based on labor economics or institutional economics (Lee, Lim, & Cho, 2022).

Moreover, the aforementioned economic studies were inclined towards *ex-post* analysis and/or grand, *macro*-dimensioned narratives; such topics included 'complement-substitute debate' of labor immigrants and native workers (Jung & Kim, 2020; Seo, 2019), the income redistribution effect of immigration from labor to capital (Kim, 2018; Choi, 2009), and examination of net policy effects (Moon, 2021; Cho & Lee, 2021). As such, a novel approach of labor immigration economics, that is, *ex-ante* anatomy on the *micro*-economic level, is needed; one such that cuts directly through 'process', rather than an 'outcome' of EPS.

**2.2** *Economic Theories*

**2.2.1** *Self-selection of Immigrants*

In 1951, in terms of labor economics, A. D. Roy provided the theoretical foundation for self-selection of labor suppliers caught in the choice between the two industries (Roy, 1951).[2] Afterwards, Borjas refined Roy model (1951) through mathematization, then extended its implications to self-selection of labor immigrants (Borjas, 1987; 1991); the key argument of Borjas' self-selection model is that the potential immigrants choose to migrate circumstances in which

$$w_1 - w_0 - M > 0 \qquad \text{[1]}$$

($w_1$: wage in host nation, $w_0$: wage in sending nation, $M$: moving/transaction costs)

is satisfied – that is, in labor immigration, the expected wage(s) of immigrants are indeed the function of their self-selection mechanism.

More specifically, in Borjas model (1987; 1991), the self-selection of labor immigrants is recognized as a dichotomy between the following two: positive and negative selection of immigrant flow. In the views of the destination country, the former attracts high-skilled immigrants; whereas the latter lures low-skilled immigrants. Positive selection takes place when host-nation's returns to skill is higher than that of sending-nation, and at the same time, the rate of returns to skill -that is, the slope of returns to skill-curve- is also higher (steeper) in the host nation. Contrastingly, negative selection occurs in the situation where host-nation's returns to skill is higher, but its rate of returns to skill is lower -that is, the returns to skill-curve is relatively flatter- that that of source country.

In short, utilizing a key parameter of 'returns to skill', Borjas successfully modelized the 'process' of labor immigration. However, Borjas' self-selection model assumes that the wages of labor suppliers are only affected by their 'universal' skill level; thus, the model overlooks the effect of aforementioned 'intranational skill-relevance' on immigrant's self-selection. In other words, in the process of immigrant flow, due to 'discount of intranational skill-relevance ($\delta$)', even an identical labor supplier could convey idiosyncratic skill levels depending on his/her residence.

---

[2] In detail, Roy model (1951) provided a methodology in which one can eliminate/control ability bias of two cohorts with respective comparative advantage in two distinct technologies.

Furthermore, Borjas' 'general' model incorporates both positive and negative selection; therefore, while the model may well-fit the immigration dynamics of traditional and pioneer host nations, latecomer countries -in which the dominant immigration flow is negatively selected- do not match the description so well. As such, pertaining to realities of short-term and low-skilled labor immigration, a 'particular solution' of latecomer destination countries, not the 'general solution' of melting-pot (or salad-bowl) nations, is of utmost necessity and significance.

**2.2.2** *Contract Theory*

Contract theory is a mathematical analytical framework into contracts, organizations, and institutions; some view it as a specialized field of jurisprudential economics- where others deem the theory as an information economics of sort (Lee, Lim, & Cho, 2022). Nonetheless, a contract theory-based modeling requires two key concepts (Bolton & Dewatripont, 2004): Individual Rationality (IR constraint) and Incentive Compatibility (IC constraint). IR denotes a condition in which a potential party will 'participate' in the contract at the very least; IR constraint requires that the utility of participation would be more than or at least equal to the reservation utility. IC refers to the utility (or profit) maximization of all parties in the contract; that is, the designer of a contract must take into account (compatibilize) the incentive structure of every potential participant in before making a proposal.

Moving on, a decomposition of contract model, that is, solving the optimization problem of each participant in the contract, utilizes an economical *rosetta stone* by the name of 'game theory'. In detail, a contract involving two parties in two -or more- periods are characterized as a 'sequential game' of a kind; accordingly, the solution to the sequential game is named the 'Subgame Perfect Nash Equilibrium (SPNE)' – with its deciphering process called the 'backward induction'.[3] In backward induction, one ascends from the final stage of sequential game (i.e. contract) to the initial starting point, thereby specifying SPNE; the aforementioned IR and IC conditions are reflected as the respective constraints to the utility-maximization problem of contract participants (or game players).

**2.3** *Skill-Relevance-Self Selection Model*

**2.3.1** *Construction of the Model*

Based on the contract process of EPS, we (Lee, Lim, & Cho, 2022) constructed the following three-staged sequential game, in which participants are the employers of EPS-implemented firm (Principal, P) and their respective employees (Agent, A): (1) P proposes a piece-rate wage contract, in which there are no fixed wages but only incentive schemes. (2) Through mediators such as Ministry of Employment and Labor, A checks and decides whether or not (s)he will participate in the labor contract. Then (3) A is stationed at host-nation EPS-firm, which by then A chooses his/her optimal effort level so as to maximize his/her utility.

The model presupposes both contract entities (P and A) to be risk-neutral. Furthermore, we assumed no asymmetric information, thus no moral hazard nor adverse selection takes place – for that the EPS-implementing firms are small-to-medium sized, and therefore P can keep track of A's effort level with relatively low monitoring costs than big firms; and for that the mediators facilitate the 'complete' information exchange between P and A in the selection process. A noteworthy feature of the game is that the labor contract P proposes to A in (1) is an ultimatum game (that is, take-it-or-leave offer); hence, in stage (2), A's strategies are only the following two – accept or decline. P and A's payoff function, or profit/utility function, are designed as **[2]-[6]**:

---

[3] Subgame Perfect Nash Equilibrium (SPNE) denotes a Nash Equilibrium in which every players of the game exercise the optimal choice in every subgame.

$$V = R(a, s_1) - W_1(a) \qquad [2]$$
$$U_1 = W_1(a) - C_1(a) - M \qquad [3]$$
$$U_0 = W_0(a) - C_0(a) \qquad [4]$$
$$R(a, s_1) = s_1 a, \qquad W_i(a) = w_i a \qquad [5]$$
$$C_i(a) = c_i a^2, \quad c_i = \frac{1}{s_i} \;\forall i, \quad \Delta w = w_1 - w_0, \quad \delta = s_0 - s_1 \qquad [6]$$
$$(c_1 > c_0 > 0, \; 0 < s_1 < s_0, \; i \in \{0,1\}, \; a > 0)$$

Where $W_i(a)$ denotes the wage A is offered in country $i$ ($i \in \{0,1\}$); $M$ denotes the moving cost, or transaction cost of immigrating from country 0 (sending-nation) to country 1 (host-nation). As mentioned above, P only offers piece-rate wage with no fixed salary; subsequently, the wage rate in country $i$ is denoted as $w_i$. When P designs her/his labor contract in regards to profit maximization, $w_0$ is exogenously determined by country 0's firms; thus, by $\Delta w = w_1 - w_0$, P's choice variable is $\Delta w$, that is, 'additional wage' – or, namely, 'wage differential' of immigration.[4] Likewise, as the revenue of employer in country 1 is only affected by the effort level of A ($a$) and A's intranational skill-relevance in country 1 ($s_1$), P's revenue is denoted by $R(a, s_1) = s_1 a$; that is, $s_0$ is irrelevant from P's revenue in country 1.

$C_i(a)$ denotes the effort cost A has to undergo in country $i$ ($i \in \{0,1\}$); where $C_i(a) = c_i a^2$ is a quadratic function -that is, a convex increasing function- with respect to $a$, in reflection of the law of increasing marginal costs. Naturally, as intranational relevance of skill ($s_i$) increases, the coefficient of effort cost ($c_i$) will be diminished; as such, $c_i$ and $s_i$ are designed as having a reciprocal relationship with each other. Meanwhile, in country 1, A's occupation of 'secondary industry' would pose potential harm to his/her body and life (since they are '3D jobs': Dirty, Difficult, and Dangerous), and A would most likely experience lingual, cultural barriers in everyday life; hence, A would encounter a higher effort cost in country 1 than in country 0 – consequently, $c_1 > c_0 > 0$. Since $c_i$ and $s_i$ are reciprocal, $c_1 > c_0 > 0$ translates to $0 < s_1 < s_0$; here, $\delta = s_0 - s_1$ is set as a parameter implicating 'discount of intranational skill-relevance' from immigration.

**2.3.2** *Solution to the Optimization Problem*

Using aforementioned backward induction method, we (Lee, Lim, & Cho, 2022) deciphered the characterized sequential game by solving P and A's optimization problems: the final stage of the contract is (3) - in which A chooses her/his optimal effort level ($a$) as a choice variable, so as to maximize his/her objective function $U_1 = W_1(a) - C_1(a) - M$ (the expected utility of immigration). A's optimization problem is given by **[7]**; since the Second Order Condition (SOC) is always negative (∵ $0 < s_1$), the First Order Condition (FOC) is indeed the solution, that is, the Incentive Compatibility (IC) constraint, to A's optimization problem – refer to **[8]** and **[9]**.

$$Max\; E(U_1) = W_1(a) - C_1(a) - M = (\Delta w + w_0)a - \frac{1}{s_1}a^2 - M \qquad [7]$$

$$FOC: \frac{\partial U_1}{\partial a} = (\Delta w + w_0) - \frac{2}{s_1}a = 0, \qquad SOC: \frac{\partial^2 U_1}{\partial a^2} = -\frac{2}{s_1} < 0 \qquad [8]$$

$$\therefore a = \left\{\frac{s_1(\Delta w + w_0)}{2}\right\} \; (IC) \qquad [9]$$

Moving backward, for A to participate in the labor contract offered by P in stage (2), A should be better off or at least indifferent by choosing to immigrate to country 1; that is, $U_1 \geq U_0$ must hold. Presumably,

---
[4] Since the employer in host-nation (country 1) cannot influence the piece-rate wage in sending-nation (country 0), the wage rate of country 0 is an exogenous variable with no endogeneity; as such, it is deemed as a constant.

Individual Rationality (IR) constraint is given by inequalities of **[10]** and **[11]**:

$$U_1 \geq U_0, \quad \therefore (\Delta w + w_0)a - \frac{1}{s_1}a^2 - M \geq w_0 a - \frac{1}{s_0}a^2 \quad \textbf{[10]}$$

$$(\Delta w + w_0)a \geq \left(\frac{1}{s_1} - \frac{1}{s_0}\right)a^2 + w_0 a + M \ (IR) \quad \textbf{[11]}$$

Finally -and initially-, in stage (1), P must decide piece-rate wage ($w_i$) that (s)he will offer to A, with respect to IC constraint of **[9]** and IR constraint of **[11]**; as such, substituting by **[11]**, **[2]** is reconstructed as **[12]**:

$$Max\ E(V) = s_1 a - w_1 a = s_1 a - (\Delta w + w_0)a \leq s_1 a - \left(\frac{1}{s_1} - \frac{1}{s_0}\right)a^2 - w_0 a - M \quad \textbf{[12]}$$

Here, for IR constraint to be an actually effective binding constraint, the constraint must hold as an equality condition; hence, **[12]**'s inequality is reorganized into the equality of **[13]**. [5]

$$Max\ E(V) = s_1 a - \left(\frac{1}{s_1} - \frac{1}{s_0}\right)a^2 - w_0 a - M \quad \textbf{[13]}$$

Likewise, IC constraint holds as an equality as well; substituting by **[9]**, **[13]** is stylized into **[14]**:

$$Max\ E(V) = s_1 \left\{\frac{s_1(\Delta w + w_0)}{2}\right\} - \left(\frac{1}{s_1} - \frac{1}{s_0}\right)\left\{\frac{s_1(\Delta w + w_0)}{2}\right\}^2 - w_0 \left\{\frac{s_1(\Delta w + w_0)}{2}\right\} - M \quad \textbf{[14]}$$

As mentioned above, since P maximizes $V$ with respect to her/his choice variable $\Delta w$, by taking the partial derivative of **[14]** with respect to $\Delta w$, we (Lee, Lim, & Cho, 2022) drew out FOC and SOC of **[14]**; by the assumption of $0 < s_1 < s_0$, SOC in **[15]** is always negative – thus, one can conclude that the FOC of **[14]**, which is given by **[16]**, is indeed the ultimate solution, that is, SPNE, to the optimization problem.

$$\frac{\partial^2 V}{\partial \Delta w^2}\left[\left\{\frac{s_1{}^2(\Delta w + w_0)}{2}\right\} - \left(\frac{1}{s_1} - \frac{1}{s_0}\right)\left\{\frac{s_1(\Delta w + w_0)}{2}\right\}^2 - w_0\left\{\frac{s_1(\Delta w + w_0)}{2}\right\} - M\right] = \frac{s_1(s_1 - s_0)}{2s_0} < 0 \quad \textbf{[15]}$$

$$s_0 = \frac{s_1(\Delta w + w_0)}{2w_0 - s_1 + \Delta w} \quad \textbf{[16]}$$

Interestingly, the only case that the parameters and variables of **[16]**, that is, the SPNE, satisfy the aforementioned presupposition of $0 < s_1 < s_0$ is that of **[17]**. [6]

$$w_0 = 0, \quad 0 < s_1 < \Delta w, \quad s_0 = \frac{s_1(\Delta w)}{\Delta w - s_1}, \quad \Delta w = \frac{s_0 s_1}{s_0 - s_1} \quad \textbf{[17]}$$

Based on inequality and equalities of **[17]** and their implications, we (Lee, Lim, & Cho, 2022) stylized the relationship between variables and parameters, that is, $\Delta w, s_0, s_1, \delta$, which are given by **[18]**-**[20]**:

$$\frac{\partial \Delta w}{\partial s_1} = \frac{s_0^2}{(s_0 - s_1)^2} > 0 \ (\because 0 < s_1 < s_0) \quad \textbf{[18]}$$

$$\frac{\partial \Delta w}{\partial s_0} = -\frac{s_1^2}{(s_0 - s_1)^2} < 0 \ (\because 0 < s_1 < s_0) \quad \textbf{[19]}$$

$$\frac{\partial \Delta w}{\partial \delta} = -\frac{s_0^2}{\delta^2} < 0 \ (\because \Delta w = \frac{s_0(s_0 - \delta)}{\delta} = \frac{s_0^2}{\delta} - s_0) \quad \textbf{[20]}$$

---

[5] This process is indirectly proof-able. In short, for that P is a reasonable economical entity maximizing his/her profit, (s)he has no incentive to choose the profit level strictly less than the maximum profit denoted in **[12]**; that is, (s)he always chooses her/his maximum profit, which in this case, is set as an equality.

[6] For **[17]**, there can be a case in which A's reservation utility ($U_0 = w_0 a - \frac{1}{s_0}a^2$) becomes a negative number, since $w_0 = 0$; however, a negative reservation utility is quite a common phenomenon in reality – industrializing nations often suppress wage increase in favor of firms, and ultimately, for economic growth (Scitovsky, 1985).

In **2.3.1**, $s_i$ was defined as 'intranational relevance of skill' in each country – accordingly, by economic implications of **[18]**-**[20]**, we (Lee, Lim, & Cho, 2022) attained the following three stylized facts: (i) as intranational skill-relevance in country 1 increases, the wage differential also increases. (ii) as intranational skill-relevance in country 0 increases, the wage differential is diminished. (iii) as the 'discount of intranational skill-relevance' by immigration is intensified, wage differential is mitigated accordingly. As mentioned above, though our former study conducted regressions controlling effects of either sending-nations (i) or host-nations (ii), we left empirical research on the stylized fact (iii) as a follow-up project, which is the very unanswered question this paper seeks to embark on.

## 3 Hypotheses

### 3.1 *Design of Main Hypothesis*

Based on stylized fact (iii), that is, from $\frac{\partial \Delta w}{\partial \delta} < 0$ in **[20]**, we propose the following hypothesis:

**H1 (Discount Hypothesis)**: *As the discount of intranational skill-relevance by immigration intensifies, the wage differential of international labor immigration will be diminished.*

To confirm **H1**, we set the following dependent variable; since the classification of what sector to include in the 'secondary industry' varied greatly on the respective countries, for the sake of convenience and consistency, the authors of this study concurred on operationalizing the dependent variable into the scope of manufacturing sector.[7]

**Dependent Variable (wage differential of immigration)**: *The wage differential of manufacturing sector between the host nation (country 1) and the sending nation (country 0)*

On the other hand, in continuance with our research (Lee, Lim, & Cho, 2022) which utilized the 'secondary industry share of GDP (%)' as the proxy variable for 'intranational skill-relevance ($s_i$)', and based on $\delta = s_0 - s_1$ in **[6]**, we operationalized the 'discount of intranational skill-relevance ($\delta$)' into the following independent variable;

**Independent Variable (discount of intranational skill-relevance)**: *The difference of secondary industry share of GDP (%) between the host nation (country 1) and the sending nation (country 0)*

### 3.2 *Hypotheses with Control Variables*

However, estimating the DV by simple nominal values would be an invalid analysis overlooking the following control variable:

**Control Variable 1**: *The discrepancy between nominal wage and real wage by purchasing power inequality, due to different price of goods in each economy*

As such, this study sought to control the inaccuracies of measurement (due to different price of goods) by reflecting the Purchasing Power Parity (PPP) index of 2017.[8] By all means, of which will be presented further down, the data of DV and IV are measures took in 2019 – consequently, there will be a 2-year long lag in controlling the inflation effects of each country. Nonetheless, it was an inevitable choice of authors to make use of the manufacturing sector wage and share of secondary industry data *in 2019*, so as to reflect the latest labor immigration dynamics – that is, 'normal state' immigration economics right before the distorted migration in exogenous shock of the COVID-19 pandemic.

---

[7] In general, constituent sectors of secondary industry include manufacturing, mining, construction, and so on; however, the criteria in which what occupations are indeed included in the sectors manifested discrepancy.

[8] The theories of international parity conditions are divided into the two: Interest Rate Parity (IRP) and the Purchasing Power Parity (PPP). The former explains exchange rate in terms of Financial Accounts (KA); whereas the latter illustrates exchange rate with respect to Current Accounts (CA). Furthermore, the former relies on the premise of 'no arbitrage by capital transfers'; contrastingly, the latter presupposes 'the Law of One Price (LOP) of the Traded Goods'. Hence, in this study, which analyzes 'manufacturing and secondary industry', that is, 'Traded Goods' industry, the PPP, rather than IRP, is more appropriate.

The caveat here is that the host nation typology, which is an exogenous parameter affected by immigration-historical context of individual nation, have the possibility to exert its influence to the relationship between IV and DV; as such, we take into account the following control variable:

**Control Variable 2**: *The effects of host nation type (traditional/pioneer/latecomer) on the dynamics between the discount of intranational skill-relevance and the wage differential by immigration*

In detail, traditional host nations in North America and Oceania may not experience the effect of discount so much, in part to their long history -and its subsequent historical institution- of immigration; on the contrary, latecomer host nations in Southern Europe and (East) Asia may manifest severe effect of the IV on DV. Pioneer countries in Western and Northern Europe would be somewhere in between; Hence, we posit the following hypothesis:

**H2 (Host Nation Typology Hypothesis)**: *The degree of H1, that is, the effect of the discount (IV) on wage differential (DV), will be high in latecomer; medium in pioneer; low in traditional type nations*.

Notably, due to its natural labor -or manpower- intensiveness, secondary industry -including manufacturing sector- inevitably has the characteristic of male-dominance. Therefore, the 'gender total' of manufacturing sector wage statistics may include female wage as an 'outlier' value, thereby distorting the aggregate value – which is the very reason that this study characterizes the following control variable.

**Control Variable 3**: *The gender ratio and gender wage gap of manufacturing sector in each nation*

Since female manufacturing wage may be an outlier, it is reasonable to expect that the conformity to **H1** be clearer in male statistics than in female statistics; thus, we postulate the following hypothesis:

**H3 (Gender Hypothesis)**: *The degree of H1, that is, the effect of the discount (IV) on wage differential (DV), will be more apparent in male cohorts than in female cohorts.*

**4 Data & Method**

**4.1** *Data Description*

We primarily used two types of data for this study. Firstly, for host nations, in account of the **CV2** and **H2**, 15 representative countries of 3 types (traditional/pioneer/latecomer) were subdivided into 5 regions -traditional (North America and Oceania), pioneer_WE (Western Europe), pioneer_NE (Northern Europe), latecomer_SE (South Europe), latecomer_EA (East/Southeast Asia)- with 3 nations in each; secondary industry share of GDP (OECD) and manufacturing wage (ILO) data were collected from the respective countries. Notably, our previous paper, in which regressions controlling effects of either sending-nations (i) or host-nations (ii) were conducted, latecomer_EA region consisted of 4 countries, that is, Korea, Japan, Hong Kong, and Singapore; however, in this study, in an attempt to substantiate immigration dynamics of 'formerly newly industrializing economies', namely, 'the Four Asian Dragons', and to strike a sample-size balance between the other regions, Japan was excluded from the dataset.[9]

---

[9] The Four Asian Dragons are the nowadays developed East Asian economies of Hong Kong, Singapore, South Korea, and Taiwan, which went through rapid industrialization from the 1960s to 1990s; back in the 1980s, where 'immigration change' started to take place, they were called 'newly industrializing economies'.

Secondly, for sending nations, we utilized secondary industry share of GDP (OECD) and manufacturing wage (ILO) data from 12 out of 16 countries of which MOU of E-9 EPS is in effect – Myanmar, Indonesia, Kyrgyzstan, and Uzbekistan were excluded from the dataset because of the following reasons. For the case of Myanmar, manufacturing wage surveys were only conducted 4 times in last 12 years (2011-2022) – moreover, ongoing internal disputes like Karen conflict (1949-present) and Rakhine conflict (2016-present) could have distorted the statistics. Likewise, Indonesia's latest manufacturing wage survey was that of 2015; In the case of Kyrgyzstan, there was only a single survey in 2020 (which was, of course, influenced by COVID-19) for the last 12 years (2011-2022). Lastly, Uzbekistan only reported their statistics in Official Estimates (OE), which are more susceptible to manipulation and tampering than data collected by international organizations (e.g. OECD, ILO) – furthermore, Uzbekistan lacked gender-specific wage data; that is, Uzbekistan only had gender total estimates, which were, unsurprisingly, not updated since 2018.

### 4.2 *Statistical Methodology*

Using Ordinary Least Squares (OLS) linear regression analysis as a statistical method, this study will provide the statistical findings in two separate ways: firstly, we will disclose the values pertaining to the statistical significance of the study, including R-squared value, t-value, p-value, and so on; secondly, we will determine the sign(s) of correlation (positive/negative/neutral) between the IV and DV in three gender parameters (total/men/women) utilizing scatter plots and their respective trendlines, whilst discerning the 5 regions and typology of the host nations. As aforementioned, our IV is the difference of secondary industry share of GDP (%) between the host nation (country 1) and the sending nation (country 0), and DV is the wage differential of manufacturing sector between the host nation (country 1) and the sending nation (country 0).

Prior to the visualization of results, separate findings from OLS regression tests with the down-scaled DV were also taken into account, but later discarded with consent of the authors. In detail, measurement unit of IV (%) was far too smaller than the measurement unit of DV (2017 PPP, in US $); hence, we implemented linear regression tests down-scaling the DV by taking natural and common logarithm. However, there were no significant, nor noteworthy improvements in R-squared values of the two tests – with R-square value improvement being the primary objective of down-scaling, or in a broader sense, data-fitting, the process has failed to indicate its efficacy; which is the very reason that the concerned findings were not acknowledged. As such, we present the following results with the aforementioned disclaimer regarding data fitting process.

## 5 Results

**Table 1. Scatter Plots in Total Gender Parameter**

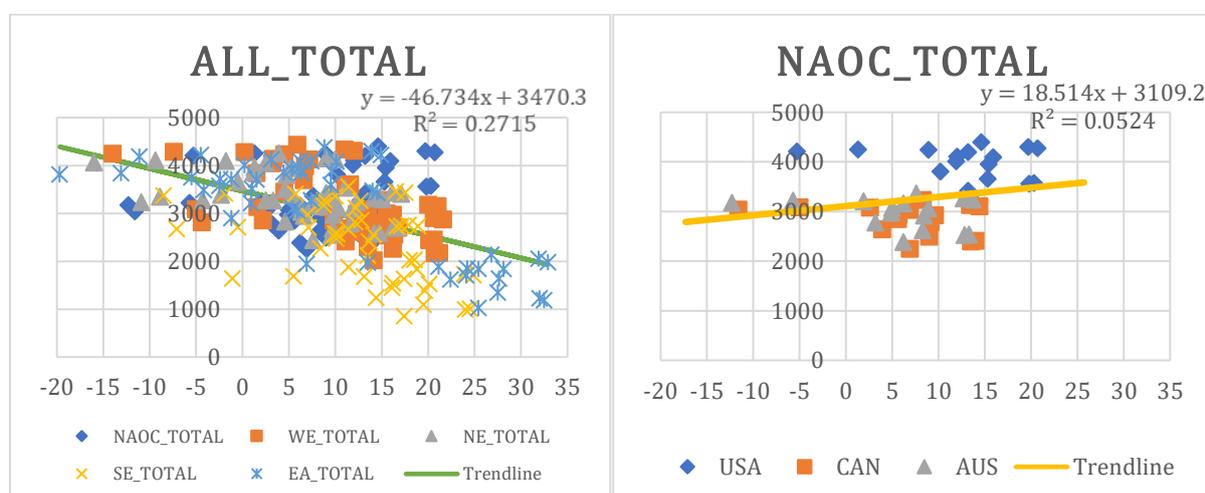

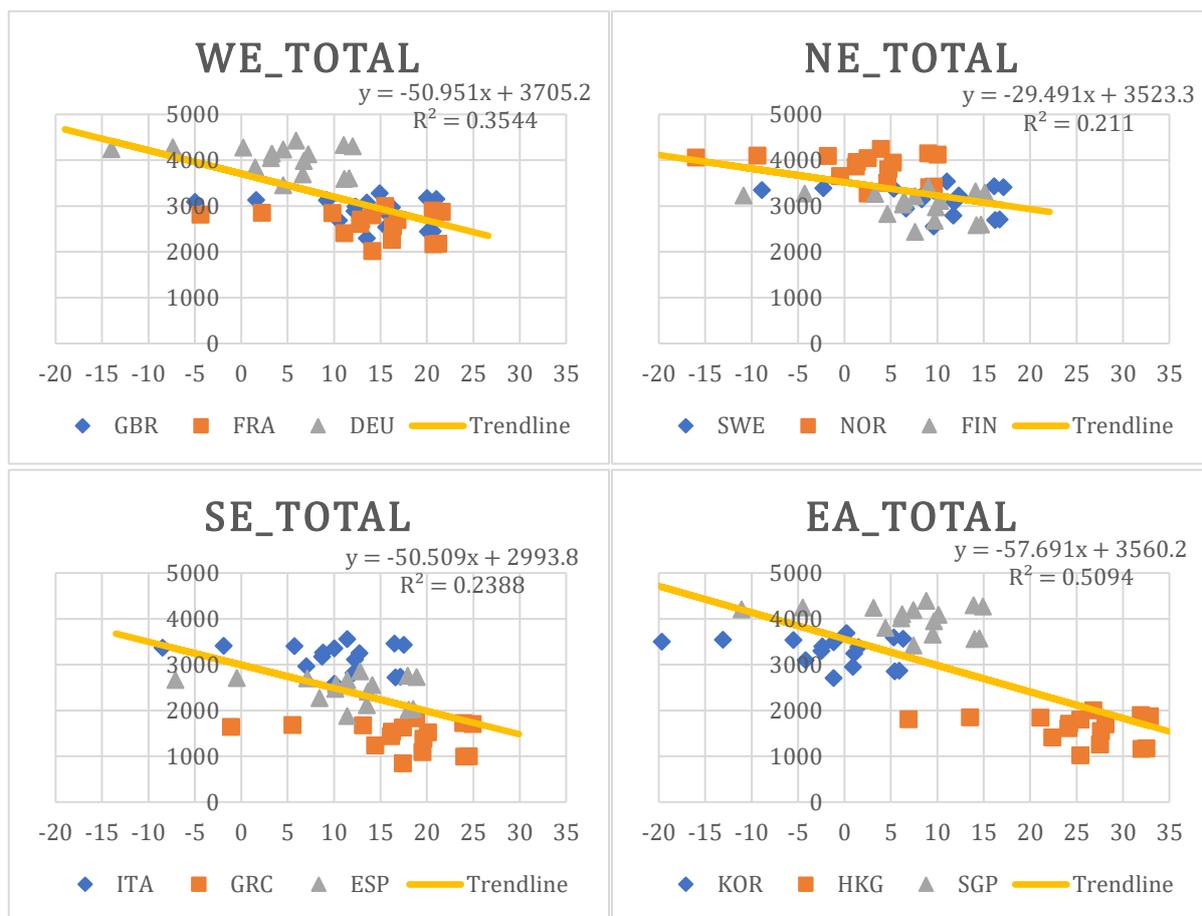

(Notation: NAOC - North America and Oceania, WE - Western Europe, NE - Northern Europe, SE - Southern Europe, EA - East Asia. Nation codes are aligned with the standard of ISO 3166-1 alpha-3.)

**Table 2. OLS Linear Regression**

| DV | Coef. | R-sq | Rev. R-sq | St. Err. | t-value | p-value | sig |
|---|---|---|---|---|---|---|---|
| ALL_TOTAL | -46.734 | 0.2715 | 0.2760 | 4.9209 | -9.5778 | 1.41e-18 | *** |
| ALL_M | -49.08 | 0.2823 | 0.2879 | 5.0213 | -9.8613 | 1.92e-19 | *** |
| ALL_F | -44.152 | 0.2845 | 0.2880 | 4.5099 | -9.8633 | 1.9e-19 | *** |
| NAOC_TOTAL | 18.514 | 0.0524 | 0.0209 | 11.4265 | 1.4084 | 0.1659 | |
| NAOC_M | 17.307 | 0.0413 | 0.0104 | 12.1077 | 1.2175 | 0.2298 | |
| NAOC_F | 11.936 | 0.025 | -0.0042 | 10.8440 | 0.8976 | 0.3742 | |
| WE_TOTAL | -50.951 | **0.3544** | 0.3448 | 10.2451 | -5.0207 | 8.58e-06 | *** |
| WE_M | -55.746 | **0.3565** | 0.3477 | 11.1513 | -5.0512 | 7.76e-06 | *** |
| WE_F | -42.767 | **0.4545** | 0.4444 | 6.9944 | -6.1476 | 1.88e-07 | *** |
| NE_TOTAL | -29.491 | 0.211 | 0.2014 | 8.4814 | -3.5501 | 0.0009 | *** |
| NE_M | -31.503 | 0.2222 | 0.2132 | 8.7622 | -3.6699 | 0.0006 | *** |
| NE_F | -31.574 | 0.2809 | 0.2771 | 7.4785 | -4.3165 | 8.59e-05 | *** |
| SE_TOTAL | -50.509 | 0.2388 | 0.2295 | 13.0268 | -3.8344 | 0.0004 | *** |
| SE_M | -54.942 | 0.2404 | 0.2315 | 14.0925 | -3.8549 | 0.0004 | *** |
| SE_F | -51.081 | **0.3151** | 0.3038 | 11.0564 | -4.5903 | 3.55e-05 | *** |
| EA_TOTAL | -57.691 | **0.5094** | 0.4938 | 8.4904 | -6.7732 | 2.21e-08 | *** |
| EA_M | -61.409 | **0.5929** | 0.5786 | 7.6157 | -8.0102 | 3.33e-10 | *** |
| EA_F | -46.911 | **0.4306** | 0.4158 | 8.0928 | -5.8084 | 6.01e-07 | *** |

*$p < 0.05$, ** $p < 0.01$, *** $p < 0.001$

For the data descripted above, Table 1 and 3 show the scatter plots and regression equations in respective gender parameters, while Table 2 displays the results of OLS linear regression. As seen in ALL_TOTAL of Table 1, the IV and DVs clearly demonstrate a negative correlation worldwide, thereby confirming **H1** (Discount Hypothesis). Furthermore, 4 regions (WE_TOTAL, NE_TOTAL, SE_TOTAL, EA_TOTAL) manifested the same negative relationship of IV and DVs, leaving out the case of NAOC_TOTAL, in which the countries of the region fall into the type of traditional host nation; interestingly, of all regions, only NAOC turned out to be statistically insignificant with the lowest R-squared values and the highest p-values (refer to Table 2).

Moreover, the findings displayed in Table 1 and 2 substantiate **H2** as well; the degree of **H1**, which can be translated to the *absolute value of negative coefficients* in regression equations above, aligns with the typology of host nations in 2.1.1 – with the average coefficient of traditional countries being +18.514 (which turned out to be statistically insignificant), pioneers -40.221, and latecomers -54.1. A possible explanation to this *actual* phenomenon is 'lock-in effects' of long-term residence in traditional host nations; that is, in the short run, migrant laborers would possess low intranational skill-relevance in country 1 (host nation), but in the long term, the discount of skill-relevance could be mitigated through On-the-Job-Training (OJT) or firm-specific technology accumulation. This inclination will be stronger in countries of which the percentage of permanent residence/naturalization seekers in immigrant population is higher than others, which is just the case of traditional host nations in NAOC.

However, the coefficients of pioneer countries were much more leaned towards the negative relationship than expected; with Western Europe (pioneer) showing a slightly stronger negative relationship than Southern Europe (latecomer), but not as much as that of East Asia (latecomer).

**Table 3. Scatter Plots in Male and Female Gender Parameters**

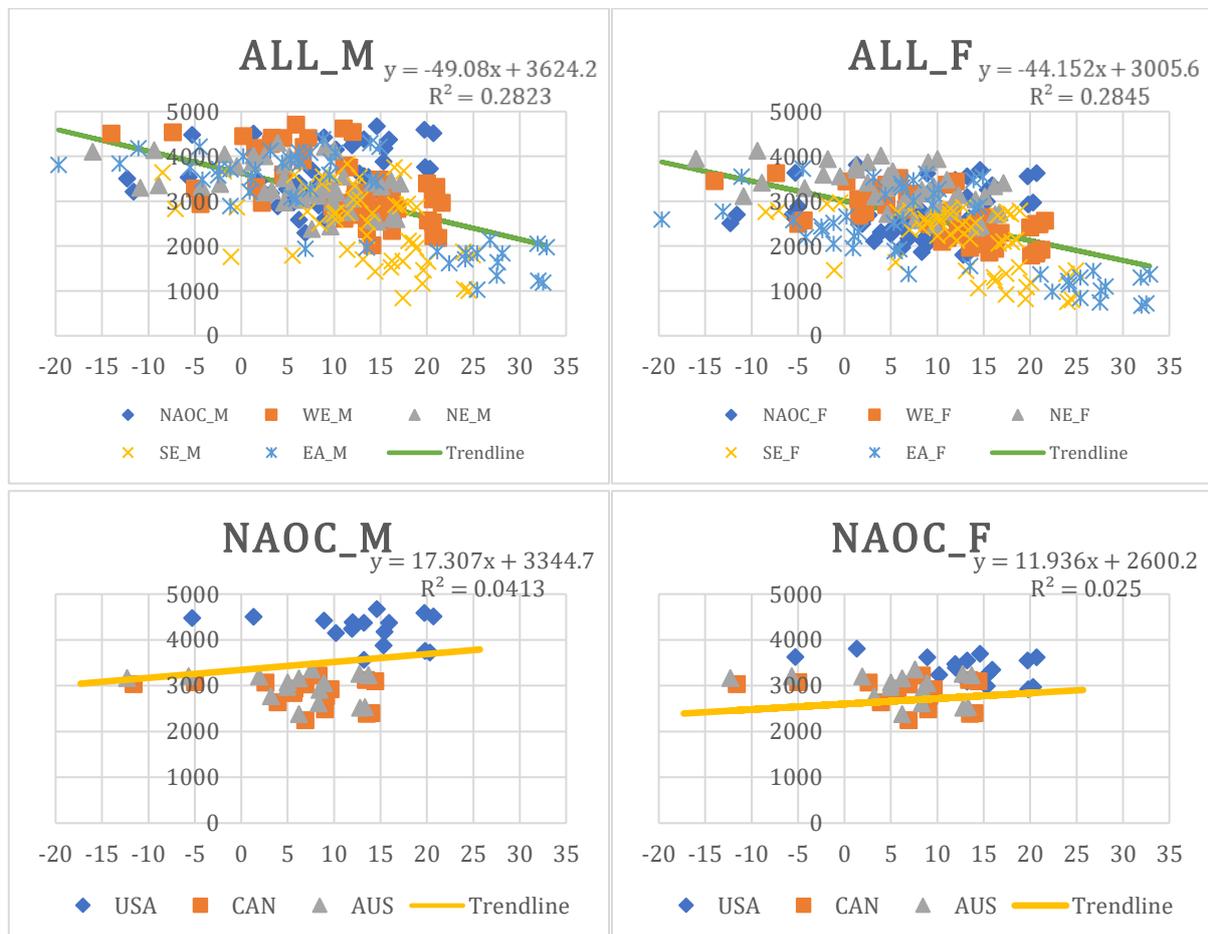

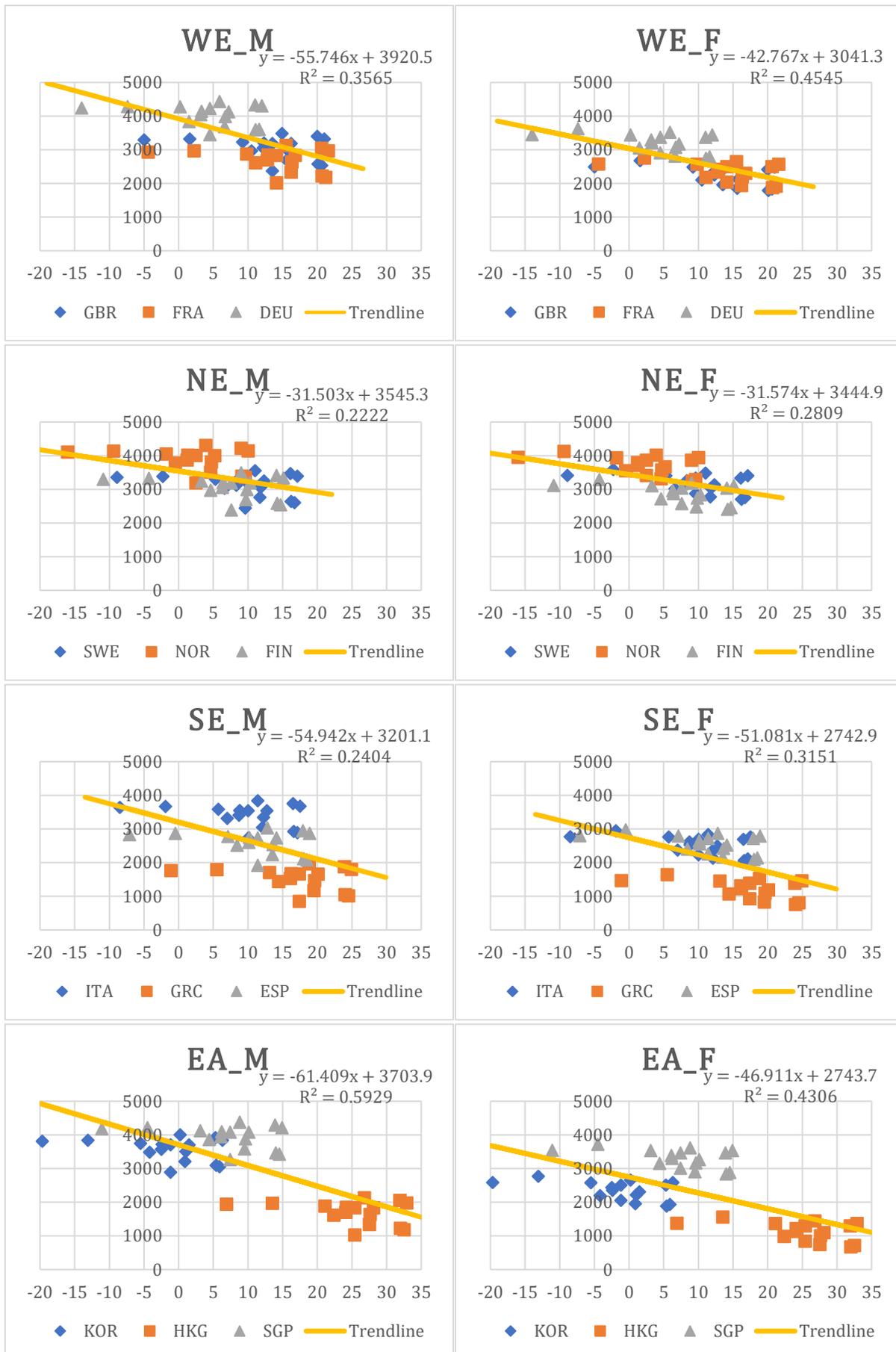

(Notation: NAOC - North America and Oceania, WE - Western Europe, NE - Northern Europe, SE - Southern Europe, EA - East Asia. Nation codes are aligned with the standard of ISO 3166-1 alpha-3.)

Excluding the statistically insignificant case of NAOC_M and NAOC_F, Table 3 generally fits together with **H3** (Gender Hypothesis), with the exception of NE (Northern Europe) countries; as shown in ALL_M and ALL_F, the degree of **H1**, that is, the *absolute value of negative coefficients* in regression equations, is higher in aggregate male parameters (-49.08) than in its female counterpart (-44.152) – 3 regions (WE, SE, EA) follow the same relationship. On the other hand, for the case of Northern Europe, the relation is slightly reversed (NE_M coef. -31.503, NE_F coef. -31.574), which can be attributed to the low domestic gender wage gap of Nordic countries (see for Lee, Lim, & Cho, 2022).

## 6 Discussion and Conclusion

This research sought to clarify and confirm the untested stylized fact of our latest immigration model (Lee, Lim, & Cho, 2022), that is, the relation between 'discount of intranational skill relevance' and 'wage differential', entailed by labor immigration. Utilizing the data meticulously selected in consideration of Employment Permit System and the typology of host nations, OLS regression results reasonably support all three hypotheses with statistical significance, thereby inductively substantiating our constructed model. Quite notably, East Asian countries, that is, the constituents of 'Four Asian Dragons', convey the greatest influence of the discount on wage differential, amongst all the other regions – pertaining to the unprecedented significance of labor immigration economics in Korea and its subsequent policy implications.

This paper contributes to existing labor immigration economics literature in three following aspects: (1) Acknowledge the previously overlooked factor of 'skill relevance discount' in labor immigration as an independent parameter, separate from the 'Moving cost' of Borjas model (1987; 1991); (2) Demonstrate the *down-to-earth* economic implications of host nation typology, thereby establishing a taxonomy *in existence*, rather than an ideal classification; (3) Seek a complementary synthesis of two grand strands of research methodology – that is, deductive mathematical modeling and inductive statistical testing.

Despite our best intentions, there remain uncovered discussions and limitations of this study. Firstly, the dominance of discount-effect in Western Europe region, which falls into the pioneer host nation category, is left unanswered; subsequently, the question of whether it is appropriate to distinguish between the pioneers and latecomers in terms of discount-effect inevitably arises. Secondly, although Korea's Employment Permit System is indeed an archetypal short-term labor immigration contract of East Asia, migrant labor systems of other countries are worth probing into. Lastly, while our model presupposed the nonexistence of imperfect information, adverse selection and moral hazard would subsist at least to some degree - which calls for follow-up studies with relaxed assumptions.

**Acknowledgements**

The authors express great gratitude for prof. Sounman Hong and prof. Taeho Eom (department chair) of Department of Public Administration at Yonsei University. We also appreciate associate prof. Sang-Hyun Kim of School of Economics at Yonsei University for his truly constructive comments.

This work is dedicated to the 37th Maeil Business Newspaper Thesis Competition in Economics. No potential conflict of interest was reported by the authors.

**Data Sources**

The authors appreciate the statistics provided by ILO Worldwide Income Database, United Nations International Migrant Stock 2019 Database, and World Bank World Development Index.

**Notes on Contributors**

*Kwonhyung Lee* is a bachelor candidate in the Department of Public Administration and School of Economics at Yonsei University. His research interests include immigration policy, immigration & labor economics, economics and jurisprudence, and so on.

*Yejin Lim* is a bachelor candidate in the International Studies major of Underwood International College at Yonsei University. Her research interests include immigration and identity, international governance, international law, and so on.

*Sunghyun Cho* is a bachelor candidate in the School of Industrial and Management Engineering at Korea University. His research interests include data analytics, healthcare operations management, operation research, and so on.

# Appendices

## Appendix 1. Host Nation Descriptive Statistics

| Host nations | ISO3166-1 alpha-3 | Secondary industry (% GDP, 2019) | 2019 manufacturing wage, total gender (2017 PPP $) | 2019 manufacturing wage, male (2017 PPP $) | 2019 manufacturing wage, female (2017 PPP $) |
|---|---|---|---|---|---|
| United States | USA | 18.3 | 4778.4 | 5081 | 4031.1 |
| Canada | CAN | 24.6[a] | 3611.2 | 3811.8 | 3101.3 |
| Australia | AUS | 25.3 | 3743.8[b] | 4105.1 | 2906.3 |
| United Kingdom | GBR | 18 | 3658.3 | 3889.4 | 2893 |
| France | FRA | 17.4 | 3381.2[c] | 3534.7[d] | 2976.2[e] |
| Germany | DEU | 27 | 4812.7 | 5112.4 | 3847.8 |
| Sweden | SWE | 21.9 | 3914.8 | 3956.6 | 3810.4 |
| Norway | NOR | 29 | 4630.1 | 4707.6 | 4346.9 |
| Finland | FIN | 23.9 | 3799.7 | 3897.1 | 3512.9 |
| Italy | ITA | 21.5 | 3942.3 | 4239.9[f] | 3164.4[g] |
| Greece | GRC | 14.1 | 2211.5 | 2361.7 | 1860.2 |
| Spain | ESP | 20.1 | 3239.6 | 3433.7 | 3194.4 |
| South Korea | KOR | 32.7 | 4073 | 4411.7 | 2992.1 |
| Hong Kong | HKG | 6.1[h] | 2383.5[i] | 2542.4[j] | 1779.7[k] |
| Singapore | SGP | 24.1 | 4778.4 | 4786 | 3946 |

---

[a] Substituted by 2018 statistics (survey not conducted in 2019).

[b] Ibid.

[c] Substituted by 2014 statistics (ADM - Déclaration Annuelle de Données Sociales data has been investigated relatively continuously from 2015 to 2020, however, the wage level tends to be tallied higher than it actually is, for that the survey takes the form of inquiring the employers about the amount of wage they pay for each employee; hence, the latest data of ES - Enquête Annuelle sur le coût de la main d'oeuvre et la Structure des salaires, which is that of 2014, is more appropriate).

[d] Ibid.

[e] Ibid.

[f] Substituted by 2013 statistics (ES – Labour-related Establishment Survey statistics has been investigated relatively continuously from 2010 to 2020, however, the data lacks gender-specific statistics; hence, the latest data of HIES – EU Statistics on Income and Living Conditions, which is that of 2013, is more appropriate).

[g] Ibid.

[h] Official Estimates by Hong Kong government, substituted by 2020 statistics (survey not conducted in 2019).

[i] Substituted by 2016 statistics (survey not conducted since 2017).

[j] Ibid.

[k] Ibid.

**Appendix 2. Sending Nation Descriptive Statistics**

| Sending nations of EPS(E-9) | ISO3166-1 alpha-3 | Secondary industry (% GDP, 2019) | 2019 manufacturing wage, total gender (2017 PPP $) | 2019 manufacturing wage, male (2017 PPP $) | 2019 manufacturing wage, female (2017 PPP $) |
|---|---|---|---|---|---|
| Philippines | PHL | 30.3 | 673.2 | 692.7 | 641.1 |
| Mongolia | MNG | 38.1 | 1220 | 1316.8 | 1104.7 |
| Sri Lanka | LKA | 27.2 | 537.5 | 658.8 | 409 |
| Vietnam | VNM | 33.72 | 830.4 | 902[a] | 774.2[b] |
| Thailand | THA | 33.6 | 1123.1 | 1198.5 | 1037.1 |
| Pakistan | PAK | 19.6 | 529.4 | 571.7 | 221.4 |
| Cambodia | KHM | 34.2 | 687.2 | 705.7 | 678 |
| China | CHN | 38.6 | 1207.8[c] | 1356.36[d] | 1059.24[e] |
| Bangladesh | BGD | 32.9 | 382.5[f] | 406[g] | 331.8[h] |
| Nepal | NPL | 13 | 571.5[i] | 601.6[j] | 401.1[k] |
| Timor-Leste | TLS | 30.2 | 770.1[l] | 838[m] | 556[n] |
| Laos | LAO | 31.5 | 583.2[o] | 703.1[p] | 479.6[q] |
| South Korea | N/A | 32.7 | 4073 | 4411.7 | 2992.1 |

---

[a] Official Estimates by Vietnam government.

[b] Ibid.

[c] Substituted by 2016 statistics (survey not conducted since 2017).

[d] Ibid.

[e] Ibid.

[f] Substituted by 2017 statistics (survey not conducted since 2018).

[g] Ibid.

[h] Ibid.

[i] Ibid.

[j] Ibid.

[k] Ibid.

[l] Substituted by 2016 statistics (survey not conducted since 2017).

[m] Substituted by 2013 statistics (survey was conducted in 2016, however, contained an extreme value).

[n] Ibid.

[o] Substituted by 2017 statistics (survey not conducted since 2018).

[p] Ibid.

[q] Ibid.

# 자기선택적 이민의 역내기술적실성 모형:
## 정형화된 사실에 대한 고용허가제도(E-9)상의 새로운 증거

이권형[*] · 임예진[†] · 조성현[‡]

<요 약>

본고는 동 저자들의 선행연구(이권형 외, 2022)에서 제시된 노동이민의 '기술적실성 자기선택' 모형에 기반, 그 함의를 확장하는 연구이다. 구체적으로, 연구자들은 당해 모형을 통해 도출되었으나 그 실체적인 증거를 미비한 다음의 정형화된 사실에 대해 경험적 검증을 의욕한다. "이민에 따른 역내기술적실성의 할인이 심화될수록, 역외노동이민의 임금 차분은 감소한다". 고용허가제(EPS)와 이민수용국 유형론을 반영하여 가설을 설립하고 데이터를 수집하였으며, 최소자승법(OLS) 선형회귀분석의 결과 모든 가설이 통계적으로 유의한 수준에서 입증된 바, 선행연구의 모형을 귀납적으로 정당화하였다. 본 연구는 기존 문헌에서 간과된 '역내기술적실성의 할인'을 Borjas 모형의 '이주 비용'과 차별화된 모수(母數)로서 새롭게 제시하며, 이민수용국 유형론의 경제 함의를 논증함으로써 그를 단순한 관념적 구분이 아닌 실체적 분류 체계로 정립하고, 노동경제학을 양분하는 연역모형 및 귀납통계 연구방법론의 상보적 합일을 모색한다.

[**주제어**: 역내기술적실성, 이민의 자기선택, 고용허가제도]

---

[*] 연세대학교 사회과학대학 행정학과, sangsankwon@yonsei.ac.kr

[†] 연세대학교 언더우드국제대학 국제학전공, yejinlim@yonsei.ac.kr

[‡] 고려대학교 공과대학 산업경영공학부, cshhello99@korea.ac.kr